\documentclass[
superscriptaddress,
preprint,
noshowpacs,
 amsmath,amssymb,
 aps,
]{revtex4-1}

\usepackage{graphicx}
\usepackage{dcolumn}
\usepackage{bm}

\usepackage{url}
\begin{document}


\title{\textbf{Surface exciton polaritons: a promising mechanism for sensing applications}}

\author{Yi Xu}
\affiliation{%
SUTD-MIT International Design Center \& Science and Math Cluster, Singapore University of Technology and Design (SUTD), 8 Somapah Road, Singapore 487372, Singapore}%
\affiliation{%
Institute of High Performance Computing, Agency for Science, Technology, and Research (A*STAR), 1 Fusionopolis Way, \#16-16 Connexis, Singapore 138632, Singapore}%

\author{Lin Wu}%
\affiliation{%
Institute of High Performance Computing, Agency for Science, Technology, and Research (A*STAR), 1 Fusionopolis Way, \#16-16 Connexis, Singapore 138632, Singapore}%

\author{L. K. Ang}%
 \email{ricky\_ang@sutd.edu.sg}
\affiliation{%
SUTD-MIT International Design Center \& Science and Math Cluster, Singapore University of Technology and Design (SUTD), 8 Somapah Road, Singapore 487372, Singapore}%


\begin{abstract}

The possibility of constructing a surface exciton polariton (SEP) based sensor at room temperature is explored. The proposed SEP sensor is based on the Kretschmann–Raether configuration of conventional surface plasmon resonance (SPR) sensor, where the metal thin film was replaced by the J-aggregate cyanine dye film. The excitation of SEP results in a strong electric field at the interface of TDBC and analyte, and exponentially decaying into the analyte, which is sensitive to the refractive index variations of analyte.  The sensitivity of 118.1818 $^\circ/\text{RIU}$ (140.4286 $^\circ/\text{RIU}$) was achieved for the proposed SEP sensor within the refractive index range 1.0-1.001 (1.33-1.36) of gaseous analyte (aqueous solutions), which is about 2 (3.5) times higher than that of conventional gold-based SPR sensor. The significant superiority of the SEP sensor in sensitivity revealing SEP as a new promising mechanism for sensing applications.      

\end{abstract}

\maketitle


\section{Introduction}
Surface exciton polariton (SEP) is an elementary excitation that results from the coupling of photons to excitons at the surface of a crystal. SEPs have been observed experimentally in various inorganic semiconductors, including ZnO \cite{lagois1976experimental,hirabayashi1982surface}, CuBr \cite{hirabayashi1976surface,hirabayashi1977surface}, CuCl \cite{hirabayashi1977surface}, and ZnSe \cite{tokura1981surface}. However, one prerequisite for the successful observation of SEPs in these media is the cryogenic temperatures, which limits the practical applications of SEPs. This is because the excitons in these inorganic semiconductors are  Wannier–Mott excitons that possesses a lower binding energy ($\sim$10 meV) than the thermal energy at room temperature (25 meV) \cite{kena2010room}. In contrast, excitons in organic crystals are Frenkel excitons with much higher binding energy ($\sim$ 1 eV) \cite{kena2010room}, which provides possibilities for the observation of SEPs in organic crystals at room temperature. In fact, SEPs in organic materials have been experimentally demonstrated at room temperture  \cite{tomioka1977surface,takatori2017surface}. For example, by replacing metal thin film in the traditional Kretschmann–Raether configuration \cite{kretschmann1968radiative} with 5,6-dichloro-2-[[5,6-dichloro-1-ethyl-3-(4-sulfobutyl)-benzimidazol-2-ylidene]-propenyl]-1-ethyl-3-(4-sulfobutyl)-benzimidazolium hydroxide (TDBC) film, Takatori $et$. $al$. \cite{takatori2017surface} have experimentally observed the SEPs under 532 nm illumination wavelength at room temperature. The observed SEPs at room temperature could pave the way for the practical applications of SEPs at room temperature.  

Optical sensing technology has been widely employed in environmental monitoring, immunoassay, medical diagnostics, and food safety \cite{estevez2012integrated,anker2008biosensing,fan2008sensitive,masson2017surface}. One of the widely used sensing mechanism is based on surface plasmon resonance (SPR) \cite{homola2008surface,mayer2011localized}, including propagating surface plasmon polariton (SPP) and localized SPR. Similar to the SEP, SPP is also an elementary excitation, which results from the coupling of plasmon to photon. A simple but classical SPR sensor    structure is the Kretschmann–Raether configuration: prism/metal/analyte, which is similar to the structure (prism/TDBC/air) supporting SEPs. With the similarities between SPPs and SEPs, then the questions emerged: is it possible to employ SEPs for sensing applications? What is the sensor performance of the SEP based sensor, better or worse than that of conventional SPR sensor? Addressing these questions is of significant value for the practical applications of SEPs, especially for sensing applications. However, as far as we know, SEP based sensors, especially operates at room temperature, have not been reported.

In this work, the SEP based gas sensor and biosensor at room temperature are theoretically investigated with the structure of prism/TDBC/analyte. It is found that the excitation of SEP is sensitive to the surrounding dielectric environment. Both the SEP gas sensor and biosensor possess significantly higher sensitivity than that of traditional gold based SPR gas sensor and biosensor. The high sensitivity associated with the proposed SEP sensor reveals the potential applications of SEPs as a new promising mechanism for sensing applications.     

\section{Design consideration and theoretical model}
\begin{figure}[thpb]
      \centering
      \includegraphics[scale=0.6]{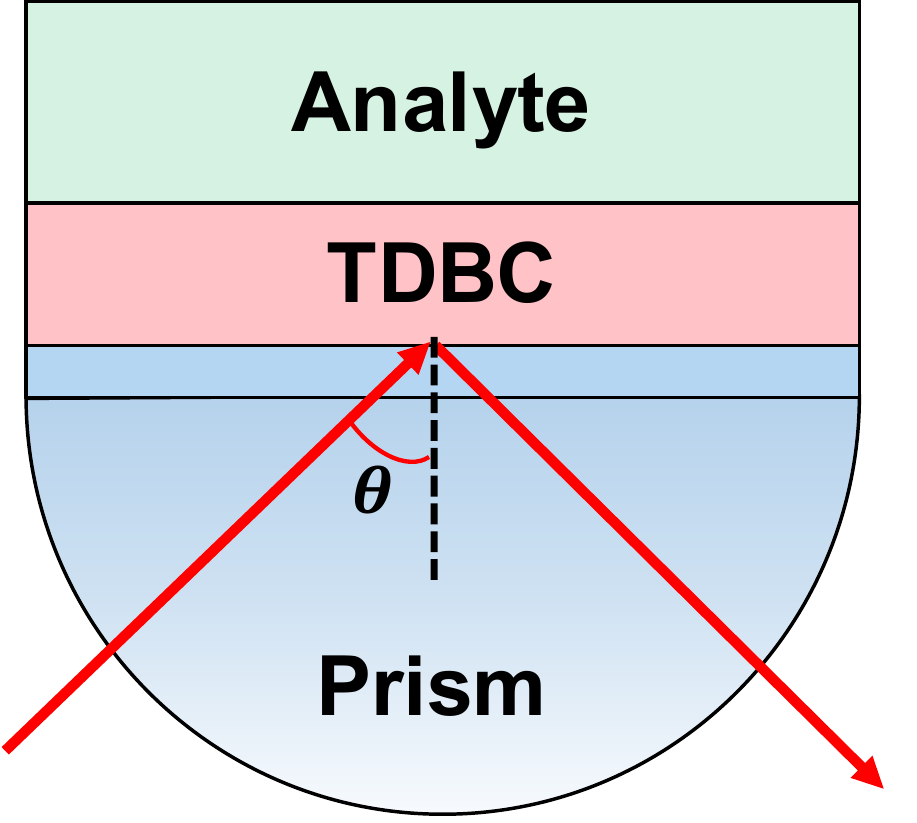}     
      \caption{Schematic diagram of proposed SEPs-based sensor.}
\label{fig1}
\end{figure}

To excite the SEPs, the traditional SPR structure, Kretschmann-Raether configuration \cite{kretschmann1968radiative}, was employed, where the surface plasmon-supporting metal film was replaced by a TDBC film, as shown in Fig. \ref{fig1}. The TDBC was deposited on a thin glass substrate, and the glass substrate was attached to a glass prism. The wavelength-dependent RI for BK7, SF11, and chalcogenide (2S2G) glass prism are, respectively, given as \cite{polyanskiy2016refractive,xu2018mos2,xu2017ultrasensitive}:

\begin{equation}
n_{\text{BK7}}=\sqrt{1+\frac{1.03961212 \lambda^2}{\lambda^2-0.00600069867} +\frac{0.231792344 \lambda^2}{\lambda^2-0.0200179144} +\frac{1.01046945 \lambda^2}{\lambda^2-103.560653} }, \\
\end{equation}
\begin{equation}
n_{\text{SF11}}=\sqrt{1+\frac{1.73759695 \lambda^2}{\lambda^2-0.013188707} +\frac{0.313747346 \lambda^2}{\lambda^2-0.0623068142} +\frac{1.89878101 \lambda^2}{\lambda^2-155.23629} },
\end{equation}
\begin{equation}
n_{\text{2S2G}}=2.24047+\frac{2.693\times10^{-2}}{\lambda^2} + \frac{9.08\times10^{-3}}{\lambda^4},
\end{equation}
where the wavelength $\lambda$ is given in $\mu$m. The dielectric constant of the TDBC film is computed through the following relation \cite{takatori2017surface}:
\begin{equation}
\varepsilon (\omega) = \sum_{j=1}^{5}\frac{\omega_{\text{p}j}^2}{\omega_{0j}^2-\omega^2-i\gamma_j\omega},
\end{equation}
where $\omega_{\text{p}1}=4340\ \text{cm}^{-1}$, $\omega_{\text{p}2}=4383\ \text{cm}^{-1}$, $\omega_{\text{p}3}=3511\ \text{cm}^{-1}$, $\omega_{\text{p}4}=11830\ \text{cm}^{-1}$, $\omega_{\text{p}5}=1621\ \text{cm}^{-1}$, $\omega_{01}=13570\ \text{cm}^{-1}$, $\omega_{02}=15330\ \text{cm}^{-1}$, $\omega_{03}=16140\ \text{cm}^{-1}$, $\omega_{04}=16960\ \text{cm}^{-1}$, $\omega_{05}=18710\ \text{cm}^{-1}$, $\gamma_1=2409\ \text{cm}^{-1}$, $\gamma_2=1352\ \text{cm}^{-1}$, $\gamma_3=565.5\ \text{cm}^{-1}$, $\gamma_4=117.3\ \text{cm}^{-1}$, $\gamma_5=561.6\ \text{cm}^{-1}$, and $\omega=1/\lambda$ is the wavenumber in $\text{cm}^{-1}$. The analyte that directly contact the TDBC film is gaseous analyte or aqueous solutions.

For a $p$-polarized light incident on the prism-coupled SEP structure, the reflection coefficient ($r_p$) is obtained as:
\begin{equation}
r_p= \frac{(\cos \beta_2 -i q_3 \sin \beta_2 /q_2 )q_1 - (q_3 \cos \beta_2 -i q_2 \sin \beta_2)}{(\cos \beta_2 -i  q_3 \sin \beta_2 /q_2)q_1 + (q_3 \cos \beta_2 -i q_2 \sin \beta_2)},
\label{rp}
\end{equation}
where 
\begin{equation}
\beta_2=\frac{2\pi d_2}{\lambda}\left(n_2^2-n_1^2\sin^2\theta_1\right)^{1/2},
\end{equation}
and 
\begin{equation}
q_j=\frac{\left(n_j^2-n_1^2\sin^2\theta_1\right)^{1/2}}{n_j^2}\ \ (j=1,2,3).
\end{equation}
Here, $n_1$, $n_2$, and $n_3$ are respectively the refractive indices of analyte, TDBC, and prism. $\lambda$ is the wavelength of incident light, and $\theta_1$ is the incident angle at the base of prism (see Fig. \ref{fig1}). Finally, the reflectance $R_p$ is given by $R_p=|r_p|^2$. 

For sensing applications, a RI variation of the analyte $\Delta n_1$ will cause a change in the reflectivity and the resonance angle ($\Delta \theta_{res}$) of SEP. Thus the angular sensitivity of a SEP sensor is defined as:
\begin{equation}
S=\frac{\Delta \theta_{res}}{\Delta n_1}.
\end{equation}
%
\begin{figure}[thpb]
      \centering
      \includegraphics[scale=0.45]{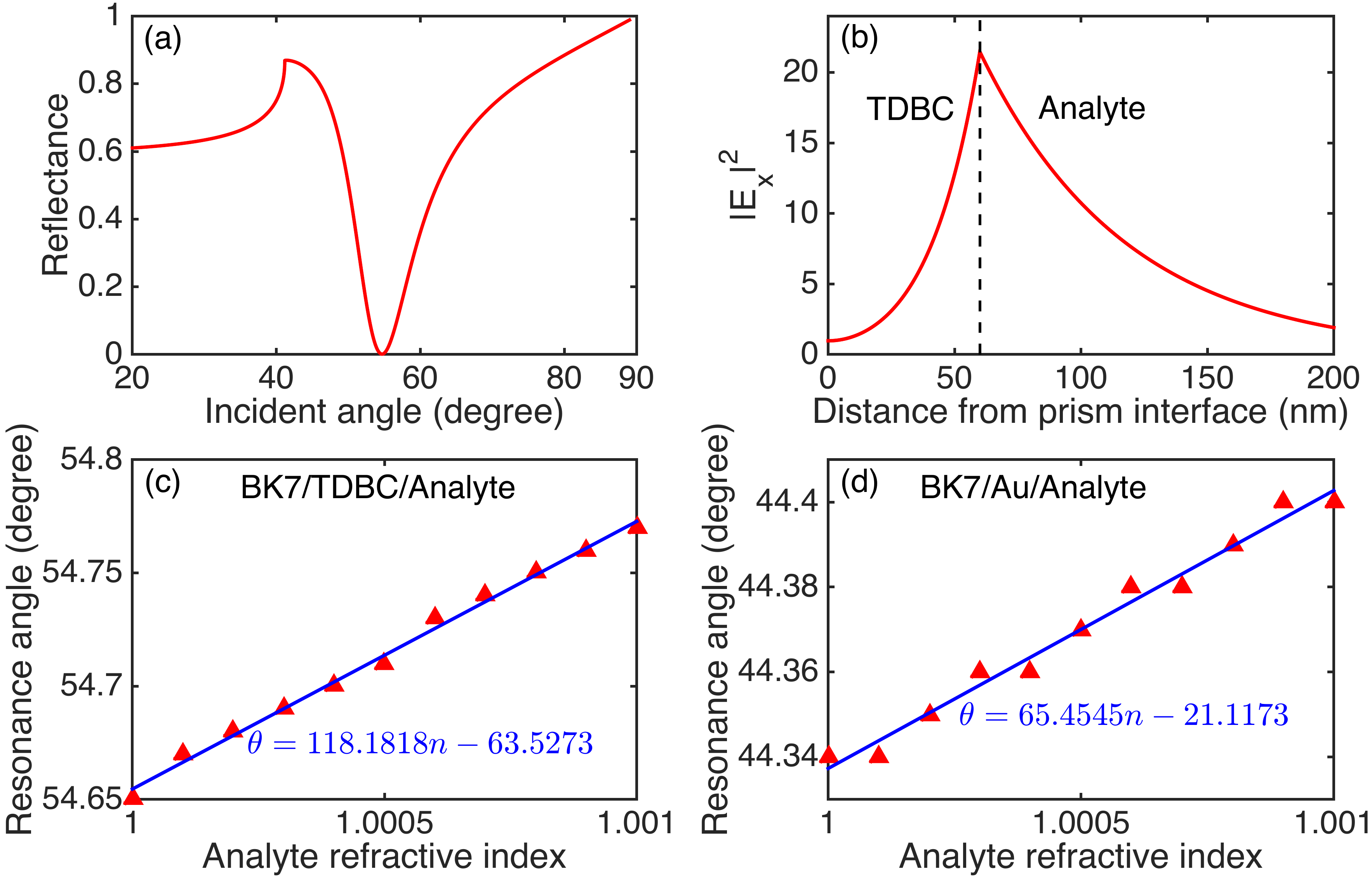}    
      \caption{SEP gas sensor with the structure of BK7/TDBC/analyte under 532 nm illumination wavelength. (a) Reflectance as a function of incident angle. (b) The density distribution of the electric field $x$-component along the TDBC and analyte regions at the resonance angle. Variation of resonance angle with the analyte RI for (c) SEP sensor (BK7/TDBC/analyte), and (d) SPR sensor (BK7/Au/analyte) . The TDBC and Au film have the same thickness, 60 nm. }
\label{fig2}
\end{figure}
\section{Results and discussion}
In this section, we would like to discuss the applications of SEP in gas sensing and biosensing. The sensor structure is shown in Fig. \ref{fig1}, where the analyte is gas (or aqueous solutions) for the gas sensor (or biosensor). The incident light is $p$-polarized with the wavelength of 532 nm.

\subsection{Gas sensing}
It has demonstrated that the TDBC film can support the SEPs within the wavelength range from 463 to 589 nm \cite{takatori2017surface}. The successful observation of SEPs has been experimentally demonstrated with the structure of BK7/60-nm-thick TDBC/air under 532 nm illumination wavelength \cite{takatori2017surface}. A resonance dip in the reflectance-angle curve was observed, as shown in Fig. \ref{fig2}(a), when the SEPs were excited. At the resonance angle, the electric field is strongest at the interface between TDBC film and analyte, and exponentially decaying into the analyte, see Fig. \ref{fig2}(b). This property of SEP is similar to that of SPR \cite{shalabney2010electromagnetic}, which is very sensitivity to the ambient changes. This in turn makes SEP a promising candidate for sensing applications. For example, this proposed SEP structure can be employed for gas sensing. The resonance angle as a function of the gas (i.e., the analyte) RI within the range from 1.0 to 1.001 is shown in the Fig. \ref{fig2}(c). It is found that the resonance angle increases with the analyte RI, and the RI sensitivity of 118.1818 $^\circ/\text{RIU}$ (RIU: refractive index unit) was achieved. For comparison, the variation of resonance angle with the analyte RI for the conventional Au-based SPR sensor (BK7/60-nm-thick Au/analyte) is plotted in Fig. \ref{fig2}(d). Similarly, the resonance angle of the SPR sensor shifts towards higher incident angle. The sensitivity of 64.4545 $^\circ/\text{RIU}$ was obtained with the SPR sensor, which is about 55.38\% of the sensitivity for the SEP based gas sensor. This revealing that SEP is a promising alternative for sensing technology with higher sensitivity than that of SPR sensor.

\begin{figure}[thpb]
      \centering
      \includegraphics[scale=0.45]{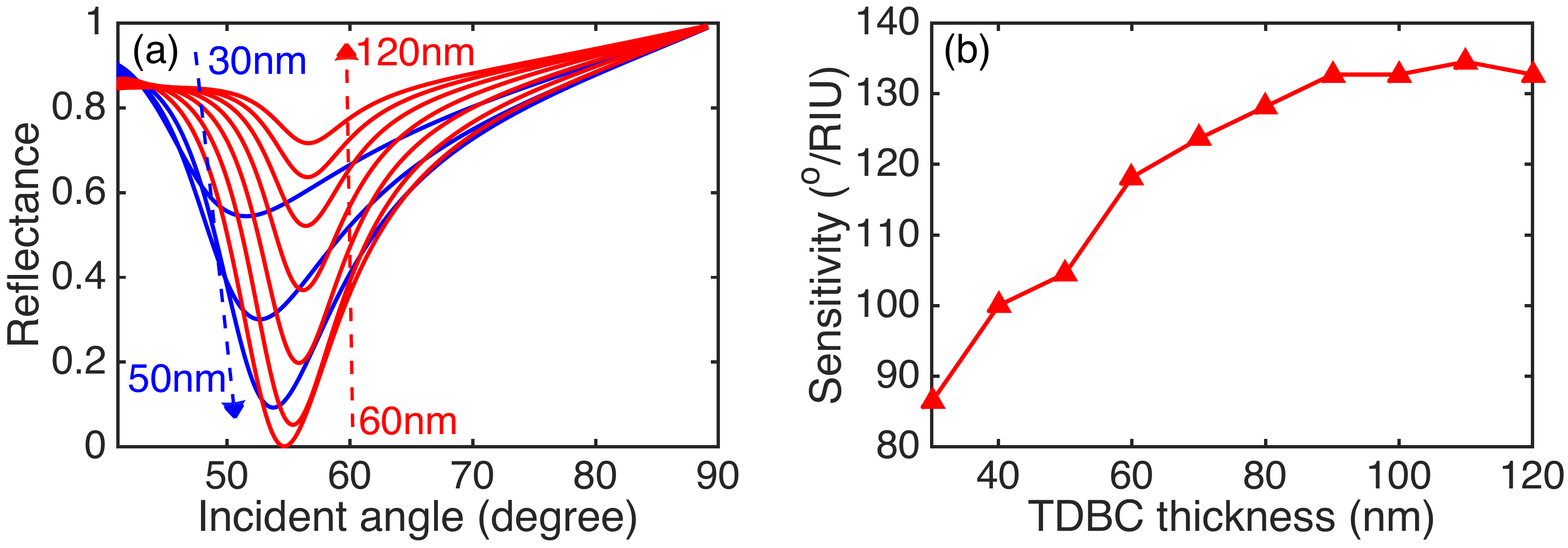}    
      \caption{(a) Reflectance as a function of incident angle for the BK7/TDBC/air structure with different thickness of TDBC film. (b) Variation of the sensitivity with the thickness of TDBC film.}
\label{fig3}
\end{figure}

The sensitivity of the proposed SEP gas sensor is strongly affected by the thickness of TDBC film. Fig. \ref{fig3}(a) illustrates the variation of the reflectance with the incident angle for SEP gas sensor with different thickness of TDBC film. It can be see from Fig. \ref{fig3}(a) that the resonance angle moves towards higher angle of incidence. The resonance depth, i.e., the minimum reflectance, is greatly dependent on the TDBC  thickness, which is related to the SEP excitation. The resonance depth first decreases and then increases with the TDBC thickness. The deepest resonance dip is found with the 60-nm-thick TDBC film, where the excitation of SEP is the strongest. For thick TDBC film, e.g., 120 nm TDBC film, it possesses a much shallow reflectance-incident angle curve, at which the incident energy absorbed by the TDBC film is insufficient to excite a strong SEP. The TDBC thickness-dependent sensitivity is shown in Fig. \ref{fig3}(b). In general, the sensitivity increases with the thickness of TDBC film in the range from 30 nm to 120 nm in spite of a plateau and one drop. Although higher TDBC film possesses greater sensitivity, its reflectance-incident angle curve is much shallow, as compared to the 60-nm-thick TDBC case. For example, the SEP gas sensor with 110 nm TDBC film exhibits the sensitivity of 134.5455 $^\circ/\text{RIU}$, while the resonance depth is 0.6366 with the ambient RI of 1.0. For the 120-nm-thick TDBC, the sensitivity is 132.7273 $^\circ/\text{RIU}$, whereas its resonance depth is 0.7167. This shallow reflectance-incident angle curve will limit the applications of SEP sensor. Taking into consideration the sensitivity and resonance depth, 80 nm-thick TDBC film can be a good choice for the SEP gas sensing that possesses a sensitivity of 128.1818 $^\circ/\text{RIU}$ and a resonance depth of 0.1978. 

\begin{figure}[thpb]
      \centering
      \includegraphics[scale=0.45]{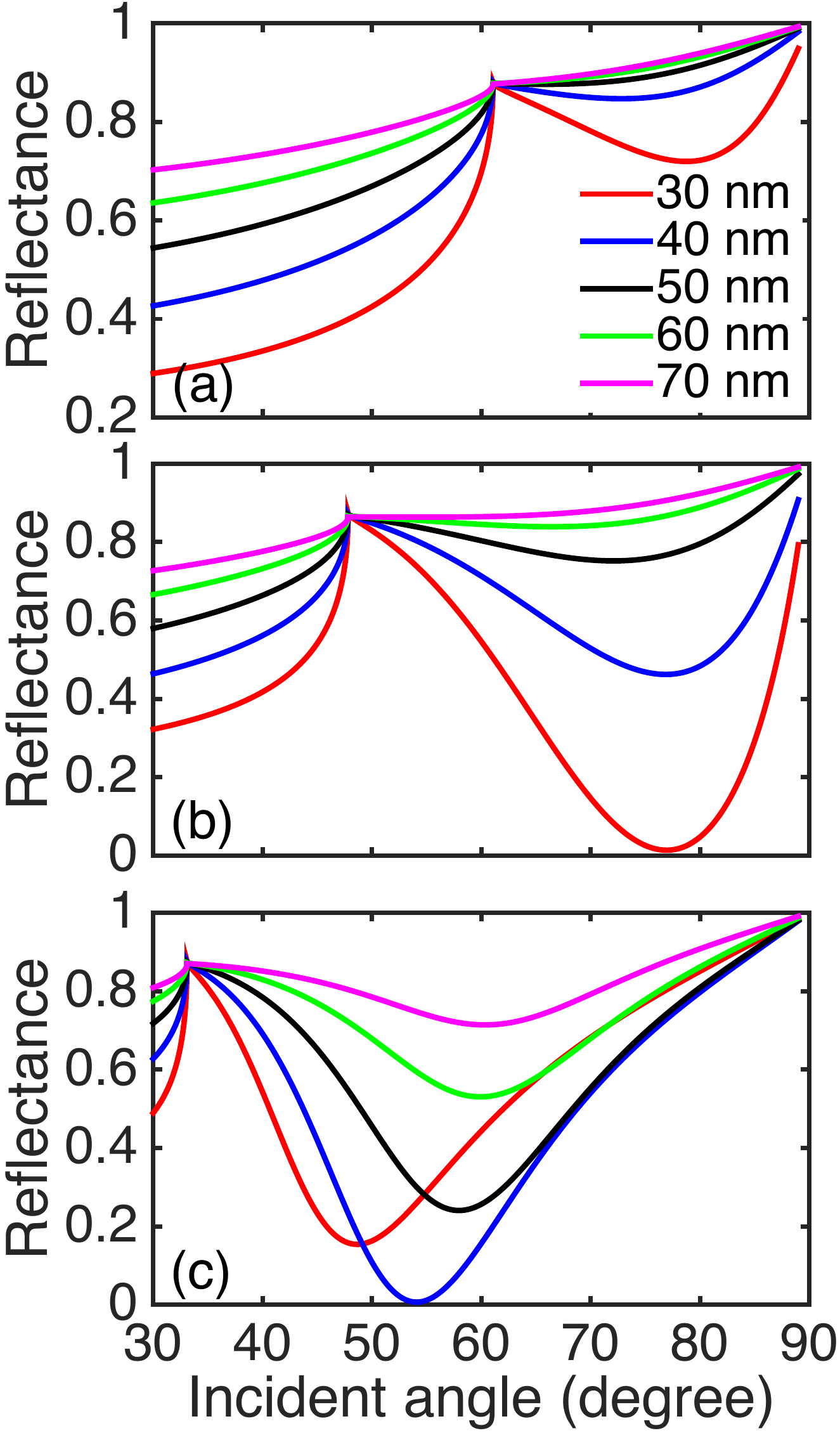}    
      \caption{Reflectance-incident angle curves for the structure of prism/TDBC/water with different thickness of TDBC film. (a) BK7, (B) SF11, (c) 2S2G.}
\label{fig4}
\end{figure}
%
\subsection{Biosensing}
We have discussed the gas sensing with the structure of BK7/TDBC/gas. But would this structure operate in a watery environment? To address this question, the reflectance-incident angle curves with the structure of prism/TDBC/water (RI$\sim$1.33) are plotted in Fig. \ref{fig4}(a). It is found that the BK7-based sensor exhibits a very shallow resonance dip and a resonance angle $\sim80^\circ$ with 30 nm TDBC film. For thicker TDBC film, it is hard to excite the SEP. The reflectance as a function of incident angle for the SF11-based sensor is shown in Fig. \ref{fig4}(b). It is found that the resonance depth increases with the thickness of TDBC film. A resonance depth of 0.0137 was obtained with 30 nm TDBC film although the resonance angle is $\sim77^\circ$. In contrast, the 2S2G-based sensor with 40 nm-thick TDBC possesses the minimum reflectance of 0.0066 at the resonance angle of 54.08$^\circ$, as shown in Fig. \ref{fig4}(c). By comparing the reflectance-incident angle curves for the three prisms, it is found that 2S2G prism is more specifically suited to biosensing. Therefore, in the following, we will focus on the structure of 2S2G/TDBC/analyte, and discuss its sensitivity. 

\begin{figure}[thpb]
      \centering
      \includegraphics[scale=0.45]{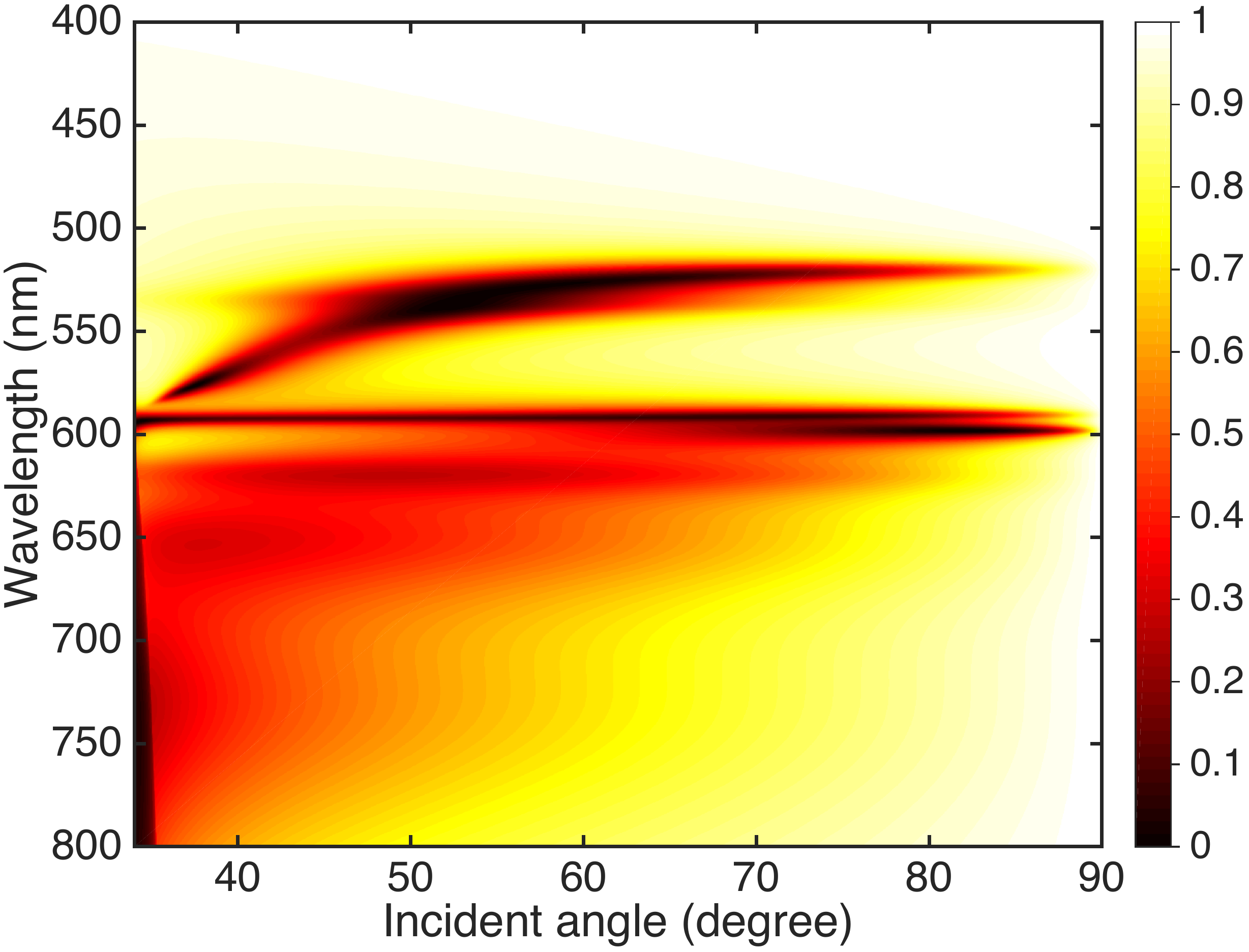}    
      \caption{The dispersion relation of an SEP with the structure of 2S2G/TDBC/water. The thickness of TDBC film is 40 nm.}
\label{fig5}
\end{figure}

The reflectance $R_p$ for the structure of 2S2G/TDBC/water as a function of the illumination wavelength and incident angle is shown in Fig. \ref{fig5}. Here, the TDBC film has a thickness of 40 nm. The locus of the  minimum reflectance in the illumination wavelength region from 515 nm to 585 nm corresponds to the dispersion relation of SEP in the structure of 2S2G/TDBC/water. This result also demonstrates the existence of SEPs within 2S2G/40 nm-thick TDBC/water under 532 nm illumination wavelength.

\begin{figure}[thpb]
      \centering
      \includegraphics[scale=0.45]{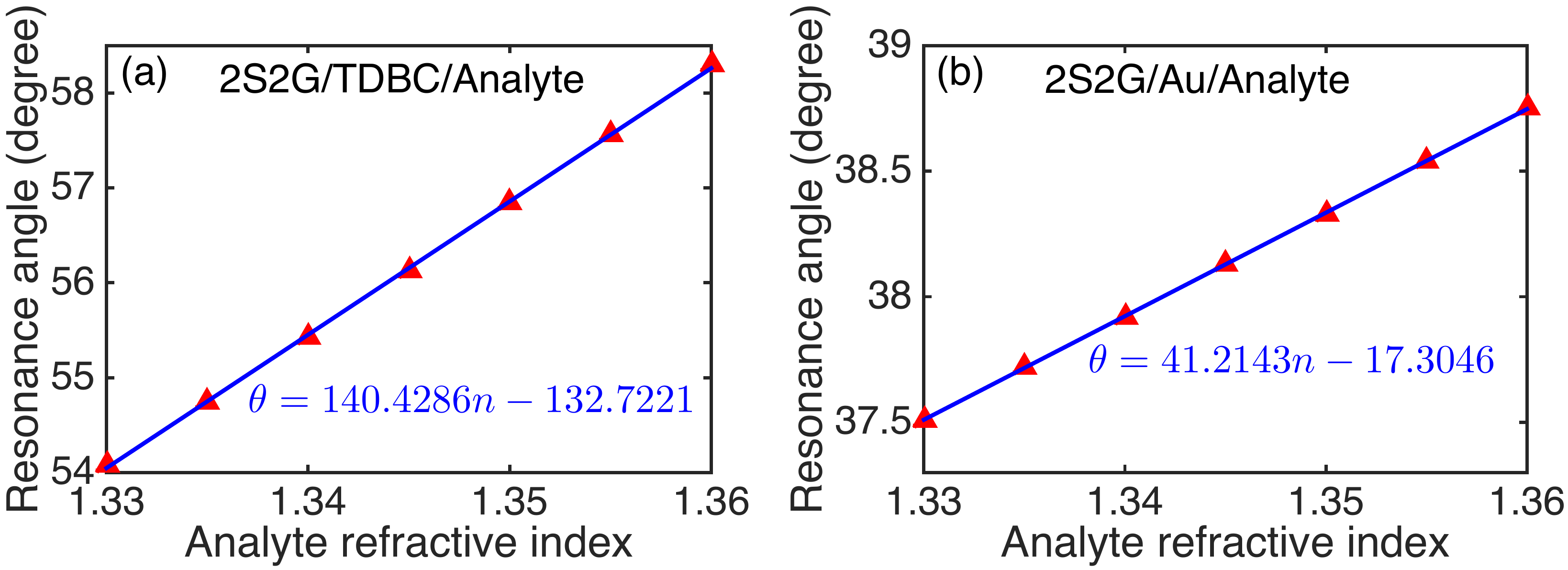}    
      \caption{Variation of resonance angle with the analyte RI for (a) SEP biosensor (2S2G/TDBC/analyte), and (b) SPR biosensor (2S2G/Au/analyte). The TDBC and Au film have the same thickness, 40 nm.}
\label{fig6}
\end{figure}

For biosensing applications, the analyte is aqueous solution, and the RI range we considered is 1.33-1.36 in this work. The variation of resonance angle with the RI of aqueous solution for SEP biosensor (2S2G/TDBC/analyte) is shown in Fig. \ref{fig6}(a). The resonance angle increases with the analyte RI, and the sensitivity of 140.4286 $^\circ/\text{RIU}$ was achieved. In contrast, the conventional SPR biosensor with 40 nm-thick Au film possesses the sensitivity of 41.2143 $^\circ/\text{RIU}$ (see Fig. \ref{fig6}(b)), which is  only 29.35\% of the sensitivity for the SEP-based biosensor. Therefore, the SEP is also an promising technology for biosensing applications.  
%

\begin{figure}[thpb]
      \centering
      \includegraphics[scale=0.5]{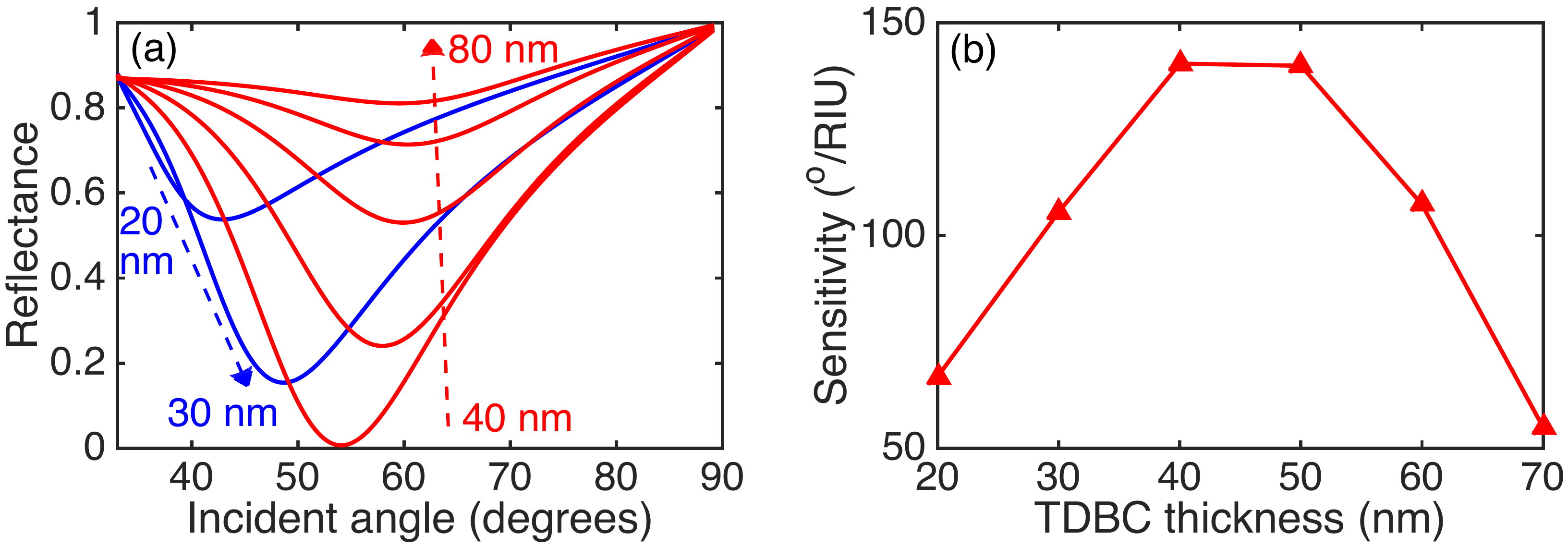}    
      \caption{(a) Reflectance as a function of incident angle for the 2S2G/TDBC/water structure with different thickness of TDBC film. (b) Variation of the sensitivity with the TDBC thickness.}
\label{fig7}
\end{figure}

The TDBC thickness-dependent reflectance-incident angle curves are plotted in Fig. \ref{fig7}(a). The resonance depth first decreases and then increases with the thickness of TDBC film. The 40 nm TDBC film shows the minimum resonance depth, while the SEP is hardly supported with the 80 nm-thick TDBC film. For the sensitivity, it first increases and then decreases with the TDBC thickness, as shown in Fig .\ref{fig7}(b), which is different from that of SEP based gas sensor (see Fig .\ref{fig3}(b)). The maximum sensitivity of 140.4286 $^\circ/\text{RIU}$ was achieved with the 40 nm TDBC film, while the 50 nm-thick TDBC based SEP sensor possesses the similar sensitivity of 139.9286 $^\circ/\text{RIU}$. For the SEP biosensor with 80 nm TDBC film, the resonance angle first increases and then decreases with the analyte RI (not shown in the paper), which is hardly to fitting with a linear function. In addition, the resonance depth of the 80 nm-thick TDBC-based SEP biosensor is greater than 0.8, which limits the practical application of the SEP sensor.   

\section{conclusion}
In this work, the possibility of the sensing applications of SEP at room temperature has been explored. At the resonance angle, the SEP possesses a strong electric field at the interface between TDBC and analyte, and exponentially decaying into the analyte. This unique property of SEP, which is sensitive to the ambient environment, provides a basis for the sensing applications of SEP. The proposed SEP sensors exhibit significant superiority in terms of sensitivity over the conventional Au based SPR sensors. For gas sensing within the RI range of 1.0-1.001, the sensitivity of 118.1818 $^\circ/\text{RIU}$ was achieved, whereas it is 65.4545 $^\circ/\text{RIU}$ for the traditional Au based SPR sensor. The SEP based biosensor possesses the RI sensitivity of 140.4286 $^\circ/\text{RIU}$ in the RI range from 1.33 to 1.36, which is more than 3 times higher than that of  Au-based SPR biosensor (41.2143 $^\circ/\text{RIU}$). With the excellent sensitivity, the proposed SEP based sensor may provide an alternative to the previous sensing technology based on SPR.

\begin{acknowledgments}
This work is partially supported by the Singapore A*STAR AME IRG (A1783c0011) and USA Air Force Office of Scientific Research (AFOSR) through the Asian Office of Aerospace Research and Development (AOARD) under Grant No. FA2386-17-1-4020.
\end{acknowledgments}


%

\end{document}